\documentclass[apj]{emulateapj}

\usepackage{graphicx,times}
\usepackage{subfigure}
\newcommand{\be}{\begin{equation}}
\usepackage{threeparttable}
\usepackage{booktabs}
\newcommand{\ee}{\end{equation}}
\newcommand{\bea}{\begin{eqnarray}}
\newcommand{\eea}{\end{eqnarray}}

\usepackage{enumerate}
\usepackage{amsmath}
\usepackage{cases}
\usepackage{longtable}
\usepackage{hyperref}
\usepackage{epstopdf}
\usepackage{amsmath,bm}
\usepackage{amssymb}
\usepackage{natbib}
\usepackage{morefloats}
\usepackage{multirow}
\usepackage{array}
\usepackage{verbatim}
\usepackage{setspace}

\begin{document}

\title{On the broad-band synchrotron spectra of pulsar wind nebulae}

\author{Siyao Xu,}
\affil{Department of Astronomy, University of Wisconsin, 475 North Charter Street, Madison, WI 53706, USA;  
Hubble Fellow; sxu93@wisc.edu}

\author{Noel Klingler, Oleg Kargaltsev,}
\affil{Department of Physics, The George Washington University, 725 21st Street NW, Washington, DC 20052, USA;
noelklingler@email.gwu.edu,
kargaltsev@email.gwu.edu}

\author{and Bing Zhang}
\affil{Department of Physics and Astronomy, University of Nevada Las Vegas, NV 89154, USA; zhang@physics.unlv.edu}

\begin{abstract}

As shown by broad-band observations, pulsar wind nebulae (PWNe) are characterized by a 
broken power-law spectrum of synchrotron emission. 
Based on the modern magnetohydrodynamic (MHD) turbulence theories, 
we investigate the 
re-acceleration of electrons in the PWN through the adiabatic stochastic acceleration (ASA), 
which arises from fundamental dynamics of MHD turbulence.  
The ASA acts to flatten the injected energy spectrum of electrons at low energies,
while synchrotron cooling results in a steep spectrum of electrons at high energies. 
Their dominance in different energy ranges leads to a flat radio spectrum ($F_\nu$) and a steep X-ray spectrum. 
Our analytical spectral shapes generally agree well with the observed synchrotron spectra of radio- and X-ray-bright PWNe. 
The spectral break corresponding to the balance between the ASA and synchrotron losses
provides a constraint on the acceleration timescale of the ASA and the magnetic field strength in the PWN.

\end{abstract}


\section{Introduction}

Poynting-flux-dominated pulsar winds (PWs) and their surrounding pulsar wind nebulae (PWNe) 
serve as a representative example for studying the mechanisms of 
energy dissipation and particle acceleration in high-energy astrophysical environments. 
Broad-band observations reveal a broken power law consisting of 
a flat radio spectrum ($F_\nu$) and a steeper X-ray spectrum as a characteristic radiation spectrum of PWNe
(e.g.,\citealt{Che05,Gae06,Rey17}).
Although the multi-wavelength radiation is believed to be synchrotron in radio and X-rays, the broad-band spectral shape is not well understood.

The stochastic nature of turbulent magnetic fields is frequently invoked to model the flat spectrum $F_\nu$ of radio emission 
(e.g., \citealt{Fle07,Ta17}).
Turbulent magnetic fields in PWNe are naturally expected and also indicated by observations 
\citep{Rey88,Yus05,Mor13,Ma16}
and simulations 
\citep{po14}.
Magnetohydrodynamic (MHD) turbulence is a highly nonlinear problem. 
Recent theoretical studies 
(\citealt{GS95,LV99}, hereafter LV99)
offer new insights into its nonlinear dynamics. 
This nonlinear character cannot be captured by a superposition of linear waves, which was traditionally assumed for modeling MHD turbulence
\citep{Giacalone_Jok1999}.
The updated understanding of fundamental properties of MHD turbulence brings 
paradigm changes in studying important physical processes, e.g., 
magnetic reconnection and dissipation
(LV99; \citealt{Zh11,Deng15,LZ18}),
particle acceleration and radiation 
\citep{PYL05,YLP08,Brunetti_Laz,Gu17,Xuc16,XZg17,Xu18,Xult},
related to high-energy astrophysical phenomena.

Alfv\'{e}n modes dominate the dynamics of MHD turbulence and carry the majority of turbulent kinetic energy
\citep{CL03}. 
The anisotropic scaling of Alfv\'{e}nic turbulence in the frame of the {\it local} magnetic field
(LV99)
is a direct consequence of its nonlinear dynamics.
Due to the turbulence anisotropy, 
the stochastic particle acceleration via gyroresonance with Alfv\'{e}nic turbulence is inefficient 
\citep{Chan00,YL02}.
For relativistic particles that undergo inefficient gyroresonance scattering,  
a new acceleration mechanism arising from the basic dynamics of Alfv\'{e}nic turbulence is applicable
\citep{BruL16}.
The stochastic dynamo stretching and reconnection shrinking of turbulent magnetic fields lead to a globally diffusive acceleration 
of particles entrained on field lines. 
Systematic energy change of particles during the trapping in individual turbulent eddies makes the acceleration highly efficient. 
\citet{XZg17,Xu18} 
term it as 
``adiabatic stochastic acceleration (ASA)"
to distinguish it from the non-adiabatic stochastic acceleration through gyroresonance scattering 
and apply it to interpreting the 
Band function spectrum commonly observed in gamma-ray bursts (GRBs)
\citep{Band93,Pre00}.
The ASA- and synchrotron cooling-dominated electron energy distributions explain well
the hard low-energy spectrum and softer high-energy spectrum of GRB prompt emission, respectively. 
As the ASA is generally applicable to particle acceleration in turbulent magnetic fields,
here we introduce the ASA and examine its effect, together with synchrotron cooling, 
on shaping the broad-band synchrotron spectra of PWNe. 
We also consider the turbulent reconnection of magnetic fields in PWs to account for 
the source of turbulence, 
as well as the injection of energetic particles for further ASA and synchrotron radiation in PWNe.

Based on the modern understanding of MHD turbulence and its interaction with energetic particles mentioned above, 
we apply our analytical model for describing the ASA and radiation loss to PWNe
and aim at providing a general explanation for their broken power-law spectra.
The structure of this paper is as follows.
In Section 2, we present our analytical model and 
elucidate the physical origin of the broken power-law spectrum in the context of a PWN. 
In Section 3, we take the Mouse PWN and some other PWNe as examples for applying
our theoretical model to the observed broad-band synchrotron spectra. 
Lastly, a discussion and a summary are given in Section 4 and Section 5, respectively.

\section{Physical origin of a broken power-law spectrum of a PWN}

\subsection{Turbulent reconnection in the PW}
\label{ssec: trec}

To facilitate the transition from the Poynting flux-dominated (i.e., high-$\sigma$) pulsar wind 
\citep{Aro02}
to the particle energy flux-dominated (i.e., low-$\sigma$) downstream of the termination shock 
\citep{Ree74,Ke84,Beg92,Bog05},
efficient dissipative processes are necessary to 
convert the magnetic energy to kinetic energy in the upstream wind or at the termination shock. 
Magnetic reconnection serves as the required dissipation mechanism
\citep{DelZ16}.
It can occur in the striped wind with alternating magnetic fields 
\citep{Cor90,Lyu01},
as well as the kink-unstable polar jet 
\citep{Beg98,Pav03,Miz11,Mig13}.

In resistive MHD theory, 
the magnetic reconnection depends on the slow resistive diffusion of magnetic fields 
\citep{Swe58,Par57}.
In the presence of turbulence, turbulent velocities dominate the magnetic field dynamics in the direction perpendicular to 
the local magnetic field
(LV99). 
The turbulent velocities act to 
increase the magnetic field gradients along the turbulent energy cascade toward smaller scales, 
and thus enhance the reconnection and determine the actual reconnection efficiency. 
In strongly magnetized turbulence, the reconnection rate is given by 
(LV99)
\begin{equation}
     v_\text{rec} \approx V_A \sqrt{\frac{L_i}{L_x}} M_A^2 ,
\end{equation}
when the injection scale of turbulence $L_i$ is smaller than the length of the current sheet $L_x$, and 
\begin{equation}
     v_\text{rec} \approx V_A \sqrt{\frac{L_x}{L_i}} M_A^2
\end{equation}
when $L_i > L_x$. 
Here $v_\text{rec}$ is the reconnection speed, $V_A$ is the Alfv\'{e}n speed, and $M_A$ is the ratio of the turbulent speed at $L_i$ 
to $V_A$. 
Without the constraint imposed by microscopic diffusion process, 
turbulent reconnection can be highly efficient
with $v_\text{rec}$ up to $V_A$. 
Numerical experiments in both non-relativistic 
\citep{KowL09}
and relativistic 
\citep{Tak15,Ea15,Zra16}
plasmas confirm that the turbulent reconnection is universal with respect to microscopic physics, e.g., resistive diffusion.

Magnetic reconnection itself acts as a source of turbulence. 
The magnetostatic free energy carried by the striped wind can be readily converted to turbulent energy 
via magnetic reconnection 
\citep{Ea15,Zra16}.
The generated turbulence in turn boosts the reconnection efficiency as discussed above 
\citep{Kow17}.
This positive feedback can result in a runaway reconnection and explosive release of magnetic energy, 
accounting for the energy source of observed $\gamma$-ray flares from e.g. the Crab Nebula
\citep{Tav11,Abd11}.
The development of both turbulence and turbulent reconnection of magnetic fields have been observed in 
numerical simulations of the pulsar striped wind, e.g., 
\citet{Zrak16}.

Moreover, magnetic reconnection can also efficiently energize particles 
\citep{DeG05,Ko11,Kow12,Del16}, 
which supplies the PWN with a power-law population of energetic electrons.
By adopting a simple treatment analogous to that for the first-order Fermi acceleration at shocks, 
\citet{DeG05}
derived a power-law index $2.5$ for the reconnection-accelerated particles.  
Beyond the termination shock, the spectrum of the accelerated electrons injected into the PWN will be affected 
by both the ASA and synchrotron cooling
in the turbulent magnetic fields downstream.

\subsection{ASA in the PWN}

Due to the magnetic reconnection in the striped wind, the PWN is expected to be fully turbulent. 
The Kolmogorov-like magnetic energy spectrum has been found in 
3D relativistic MHD simulations of PWNe, e.g., 
\citet{po14}, 
which is a typical spectral form of turbulent magnetic fields in strong MHD turbulence 
\citep{MG01,CL03}.

In strong MHD turbulence with increased turbulence anisotropy toward smaller scales, the turbulent eddies at the resonant scale 
of gyroresonance are highly 
elongated along the magnetic field. 
With the perpendicular scale of the turbulent eddies much smaller than the parallel resonant scale,
particles interact with many uncorrelated turbulent eddies within one gyro orbit, 
and thus the gyroresonance scattering is inefficient 
\citep{Chan00,YL02}.
In the absence of efficient scattering, 
relativistic particles can slide along magnetic field lines. 
Obviously, the magnetic field variation is slow with respect to the motion of particles.  
Within the traps of turbulent eddies,
the second (longitudinal) adiabatic invariant of the particles bouncing between the ``mirror points" applies. 
As there are statistically distributed reconnection and dynamo regions in strong MHD turbulence
\citep{XL16}, 
after cycles of acceleration in turbulent reconnection regions with shrinking field lines and 
cycles of deceleration in turbulent dynamo regions with stretched field lines, 
the particles have a diffusive energy gain
\citep{BruL16}.
We term this acceleration process as the ``adiabatic stochastic acceleration (ASA)"
\citep{XZg17,Xu18}.

The acceleration rate of the ASA is  
\begin{equation}\label{eq: a2}
    a_A = \xi \frac{u_\text{tur}}{l_\text{tur}},
\end{equation}
where $l_\text{tur}$ is the characteristic scale of turbulent magnetic fields, $u_\text{tur}$ is the turbulent speed at $l_\text{tur}$, 
and their ratio gives the corresponding turbulent eddy-turnover time $\tau_\text{tur}$.
$\xi$ represents the cumulative fractional energy change during $\tau_\text{tur}$. 
As relativistic particles undergo the first-order Fermi process within individual eddies, 
given the number of bounces $\sim c/u_\text{tur}$ and the energy change per bounce $\sim u_\text{tur}/c$, 
where $c$ is the speed of light, 
we can easily see that $\xi$ is of order unity. 
Thus the ASA is a very efficient stochastic acceleration mechanism compared to the 
second-order Fermi acceleration associated with gyroresonance scattering.

\subsection{Synchrotron spectrum under the effects of ASA and synchrotron cooling}
\label{ssec: case}

The stochastic acceleration generally leads to energy diffusion and a hard energy spectrum of particles. 
We consider a steady injection of a power-law distribution of electrons
originating from the magnetic reconnection sites in the PW (Section \ref{ssec: trec}), 
\begin{equation}
 Q(E) = C E^{-p},  ~~        E_l  < E < E_u.
\end{equation}
At a given reference energy, $C$ is the electron number per unit energy per unit time,
$p$ is the power-law index, 
and $E_l$ and $E_u$ are the lower and upper energy limits, respectively.
Due to the ASA, the energy distribution of electrons spreads out in energy space following 
\citep{Mel69}
\begin{equation}\label{eq: melsp}
      E \sim \exp(\pm2\sqrt{a_A t}).
\end{equation}
The injected energy spectrum evolves with time $t$
and approaches a universal form
\citep{Xu18}
\begin{equation}\label{eq: fism}
\begin{aligned}
  &    N(E,t)   
     =  \frac{C (E_u^{-p+1} - E_l^{-p+1}) \sqrt{ t} }{(-p+1)\sqrt{\pi a_A}   } 
         E^{-1} \exp \Big(-\frac{E}{E_\text{cf}}\Big)  ,\\
  &  ~~~~~~~~~~~~~~~~~~~~~~~~~~~~~~~~~~~~~~~~~~~~~~~~~~~~~~~~~~~~~~~~      E_m  < E < E_\text{cf}.
\end{aligned}
\end{equation}
The resulting energy spectrum has a minimum energy $E_m$. 
It is smaller than $E_l$ as the 
energy spectrum under the effect of ASA 
tends to have a larger energy spread than that of the injected electrons (Eq. \eqref{eq: melsp}).
$E_\text{cf}$ is the cutoff energy of the ASA and has the expression 
\begin{equation}\label{eq: cutofe}
   E_\text{cf} = \frac{a_A}{\beta},
\end{equation}
corresponding to the equalization between $a_A$ and synchrotron cooling rate. 
Here the parameter $\beta$ is given by
\begin{equation}\label{eq: betp}
    \beta = \frac{P_\text{syn} }{E^2}  
               = \frac{\sigma_T c B^2 }{ 6 \pi (m_e c^2)^2},
\end{equation}
where $P_\text{syn}$ is the power of synchrotron radiation, 
$B$ is the magnetic field strength, 
$\sigma_T$ is the Thomson cross section, 
and $m_e$ is the electron rest mass. 
Different from a steep energy spectrum of electrons, for which the minimum energy is the characteristic electron energy, 
for a hard electron spectrum in Eq. \eqref{eq: fism}, the upper cutoff energy $E_\text{cf}$ dominates the integral,
\begin{equation}
    \int_{E_m}^{E_\text{cf}} E N(E) dE 
    \approx     \frac{C (E_u^{-p+1} - E_l^{-p+1}) \sqrt{ t} }{(-p+1)\sqrt{\pi a_A}   } 
    E_\text{cf}  
     = \epsilon \dot{E}t  ,
\end{equation}
where we assume $E_u < E_\text{cf}$ and $E_m \ll E_\text{cf}$, $\dot{E}$ is the spin-down power of the pulsar, 
and $\epsilon$ is the fraction of the spin-down energy converted to the particle energy. 
To confront the above relation to observations, a detailed modeling of the particle acceleration in the PW is needed, 
which is beyond the scope of the current paper.

We next consider that the injected electrons have sufficiently high energies with $E_\text{cf} < E_u$.
The hard electron energy distribution below $E_\text{cf}$ is attributed to the ASA, 
while the distribution at energies above $E_\text{cf}$ is governed by synchrotron cooling effect. 
By defining the critical cooling energy 
\citep{Sar98}
\begin{equation}\label{eq: eneco}
     E_c = P_\text{syn} t  = \frac{1}{\beta t},
\end{equation}
we see that the injected electrons with $E > E_c> E_\text{cf}$ cool down to $E_c$ within time t.
But the electrons with $E < E_\text{cf}$ do not cool because of the ASA,
irrespective of the value of $E_c$.
It means that the overall system is always in slow cooling regime.

Given $E_m $ and $E_u$ as the lower and upper energy limits of the entire electron energy spectrum, depending on the relation 
between $E_l$, $E_\text{cf}$, and $E_c$, 
the electron spectrum at $E > E_\text{cf}$ can exhibit different forms
\citep{Xu18}.

{\it Case (i)}: $ E_c < E_l < E_\text{cf} $

\begin{figure*}[htbp]
\centering   

\subfigure[{\it Case (i)}]{
   \includegraphics[width=8.5cm]{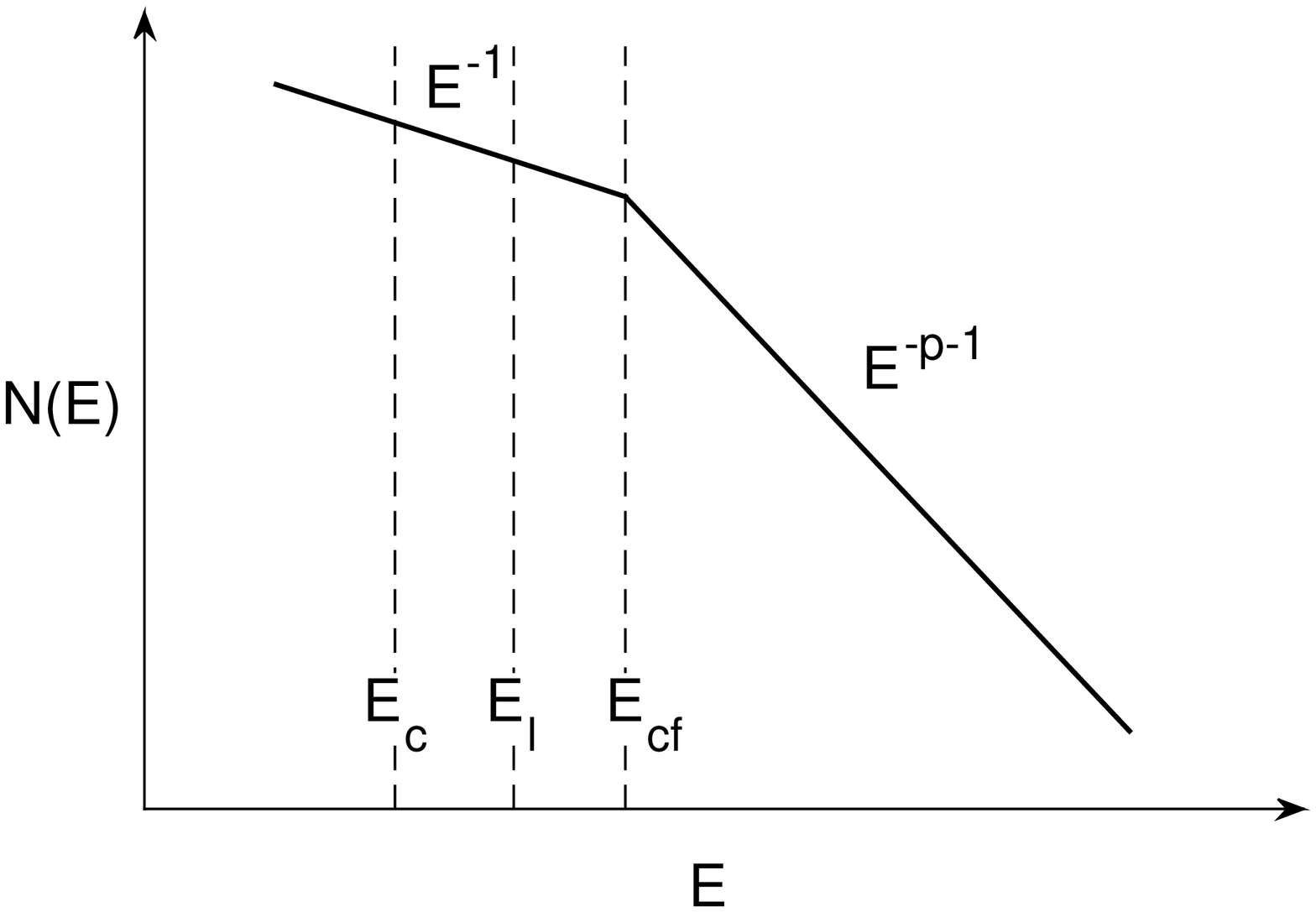}\label{fig: fac1}}
\subfigure[{\it Case (i)}]{
   \includegraphics[width=8.5cm]{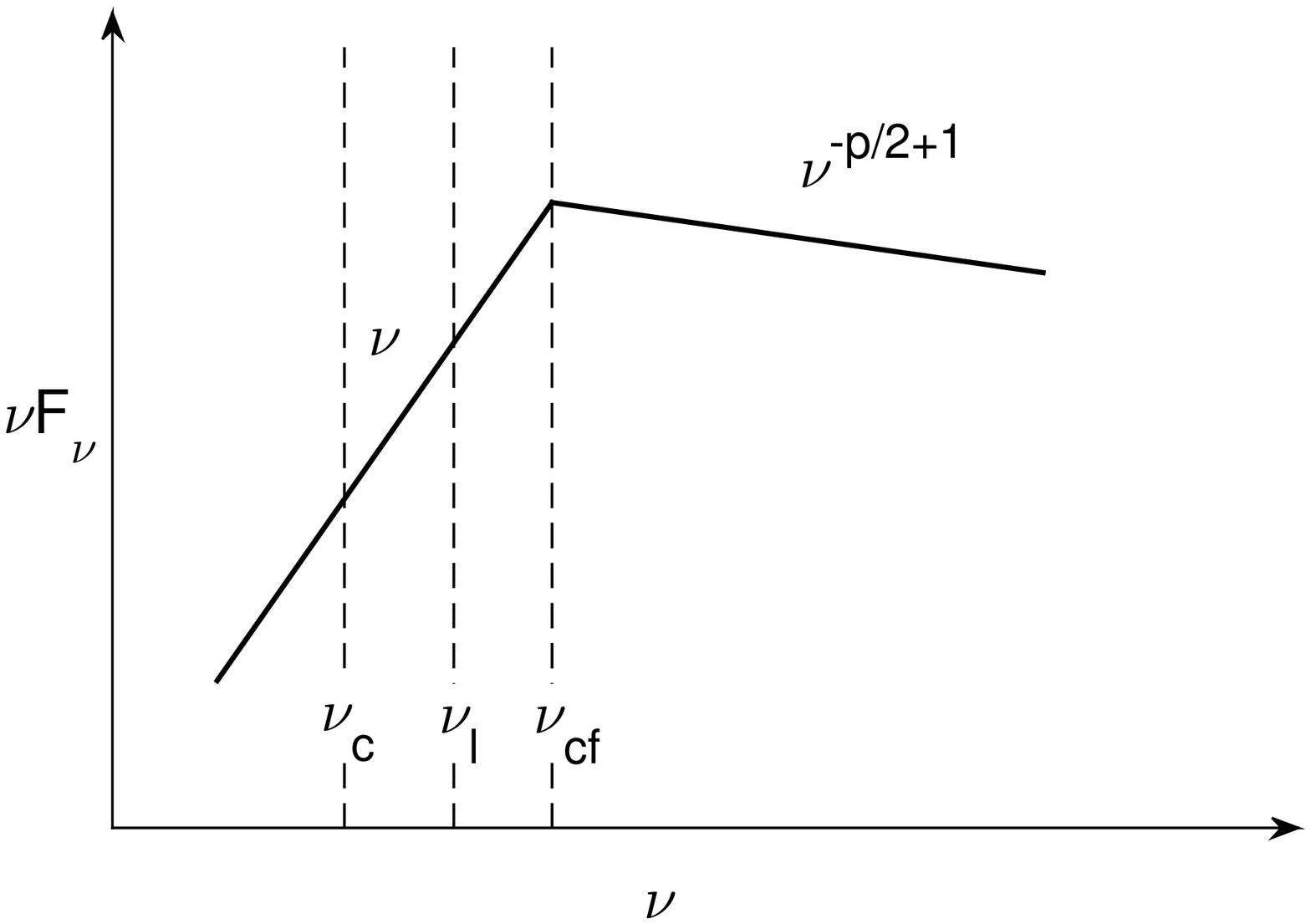}\label{fig: fac1s}}
   
\subfigure[{\it Case (ii)}]{
   \includegraphics[width=8.5cm]{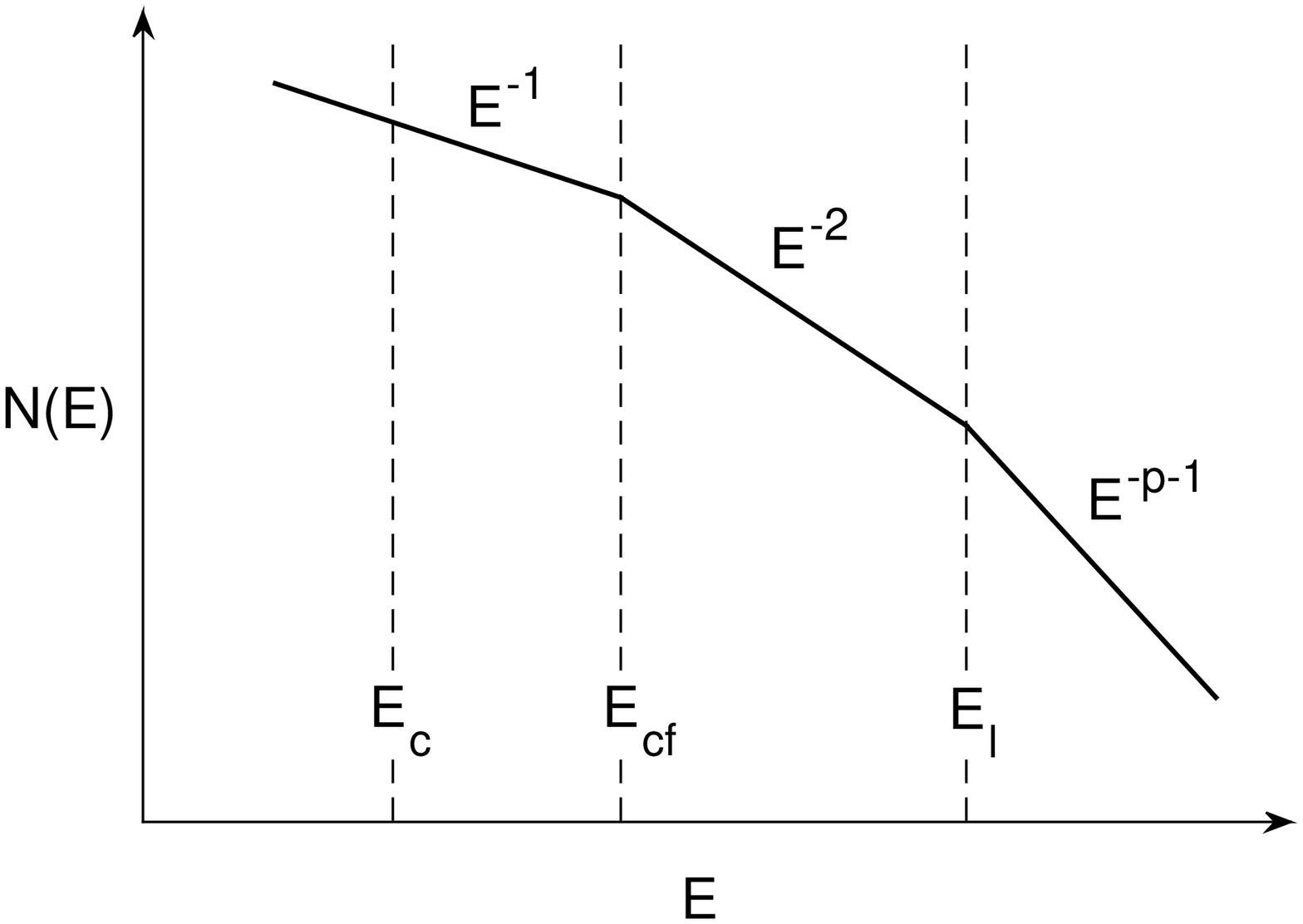}\label{fig: fac2}}
\subfigure[{\it Case (ii)}]{
   \includegraphics[width=8.5cm]{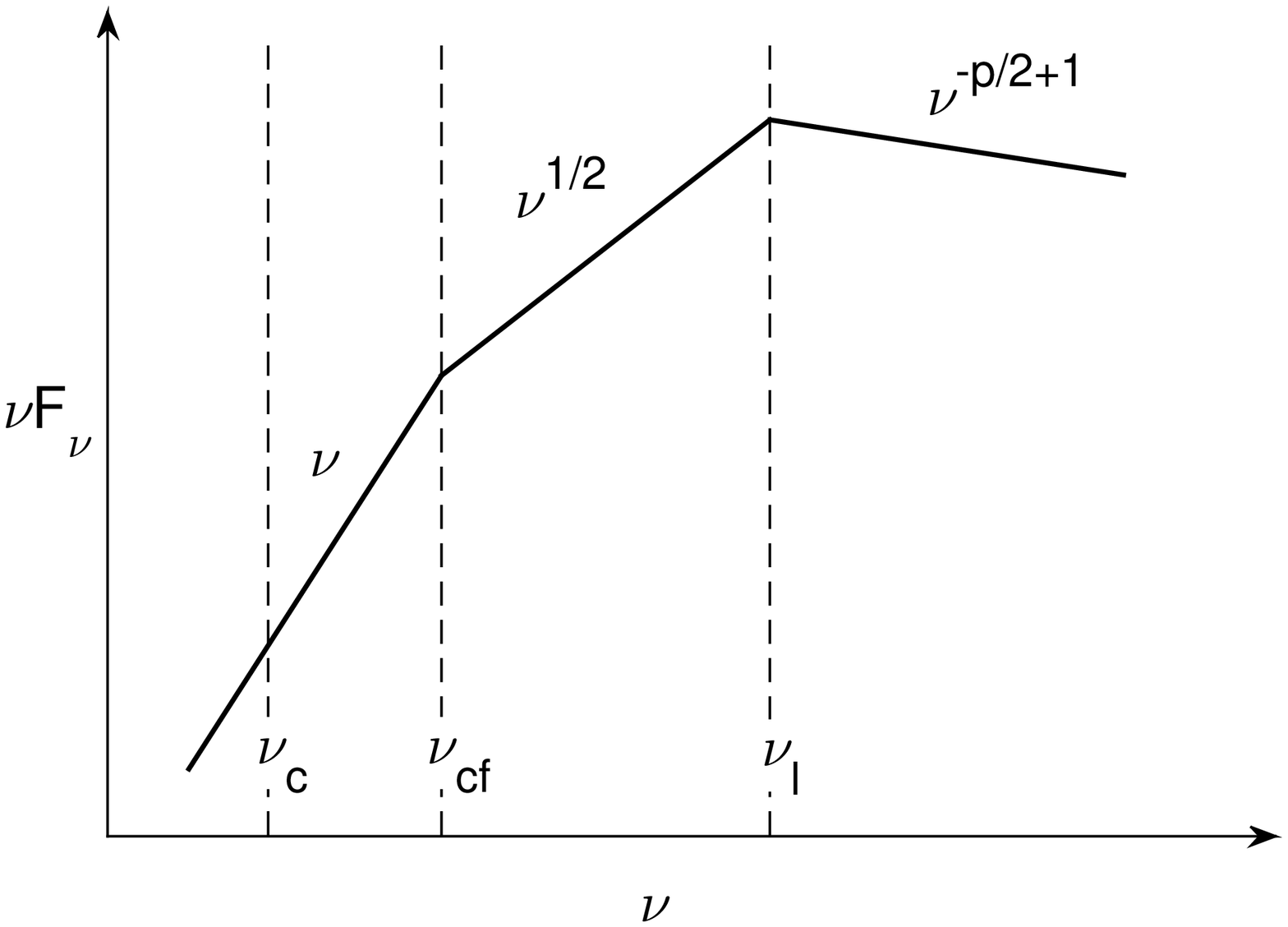}\label{fig: fac2s}}
   
\caption{The electron energy spectrum ((a) and (c)) under the effects of ASA and synchrotron cooling and 
the corresponding synchrotron spectrum ((b) and (d)) in {\it Case (i)} and {\it Case (ii)}. }
\label{fig: fast}
\end{figure*}

In the energy range above $E_\text{cf}$, the cooled electrons follow a steep spectrum 
\citep{Kar62}
\begin{equation} \label{eq: cols}
     N(E) =  \frac{CE^{-(p+1)}}{\beta (p-1)}.
\end{equation}
Combining Eqs. \eqref{eq: fism} and \eqref{eq: cols}, we illustrate the asymptotic behavior of the electron spectrum over 
a broad energy range in Fig. \ref{fig: fac1}. 
The corresponding synchrotron spectrum is 
\begin{subnumcases}
 {\nu F_\nu=\label{eq: fcc1}}
\nu F_{\nu,\text{max}} \propto \nu ,~~~~~~~~~~~~~~~~~~~~~~~~~~~~ \nu <\nu_\text{cf}, \\
\nu F_{\nu,\text{max}} \Big(\frac{\nu}{\nu_\text{cf}}\Big)^{-\frac{p}{2}} \propto \nu^{-\frac{p}{2}+1},~~~ \nu > \nu_\text{cf} ,
\end{subnumcases}
as displayed in Fig. \ref{fig: fac1s}, 
where $\nu_\text{cf}$ is the frequency related to $E_\text{cf}$, 
$F_\nu$ is the flux density at frequency $\nu$,
and $F_{\nu,\text{max}}$ is the maximum value of $F_\nu$.

{\it Case (ii)}: $ E_c < E_\text{cf} < E_l$

In the energy range above $E_l$, the same steep electron spectrum as in Eq. \eqref{eq: cols} applies. 
Moreover, driven by the cooling effect, the injected electron energy distribution extends down to lower energies. 
In the energy range $E_\text{cf} < E < E_l$, 
the electron spectrum is $N(E) \sim E^{-2}$, with the injection and radiation losses of electrons in equilibrium 
\citep{Mel69}.
Accordingly, the entire electron spectrum contains three segments, as shown in Fig. \ref{fig: fac2}. 
The corresponding synchrotron spectrum is (see Fig. \ref{fig: fac2s}),
\begin{subnumcases}
 {\nu F_\nu=\label{eq: fcc2}}
\nu F_{\nu,\text{max}} \propto \nu ,~~~~~~~~~~~~~~~~~~~~~~~~~~~~~~ \nu <\nu_\text{cf}, \\
\nu F_{\nu,\text{max}} \Big(\frac{\nu}{\nu_\text{cf}}\Big)^{-\frac{1}{2}} \propto \nu^\frac{1}{2},~~~\nu_\text{cf} < \nu < \nu_l, \\
\nu F_{\nu,\text{max}} \Big(\frac{\nu_l}{\nu_\text{cf}}\Big)^{-\frac{1}{2}} \Big(\frac{\nu}{\nu_l}\Big)^{-\frac{p}{2}} \propto \nu^{-\frac{p}{2}+1}, \nonumber\\
~~~~~~~~~~~~~~~~~~~~~~~~~~~~~~~~~~~~~~~~~~~~~~~~~~~~\nu > \nu_l,
\end{subnumcases}
where $\nu_l$ is related to $E_l$.

{\it Case (iii)}: $ E_l < E_c < E_\text{cf} $

\begin{figure*}[htbp]
\centering   

\subfigure[{\it Case (iii)}]{
   \includegraphics[width=8.5cm]{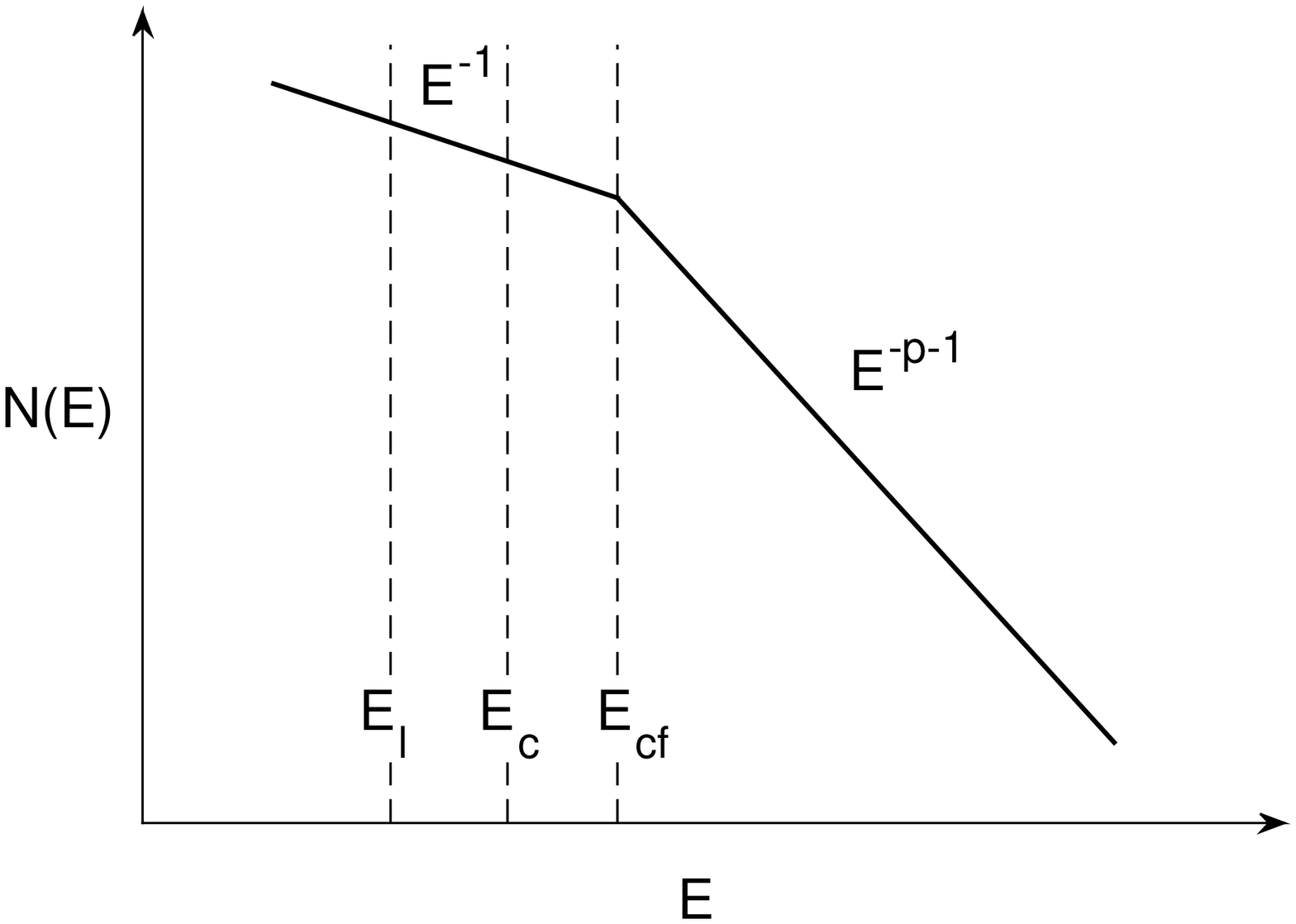}\label{fig: slc1}}
\subfigure[{\it Case (iii)}]{
   \includegraphics[width=8.5cm]{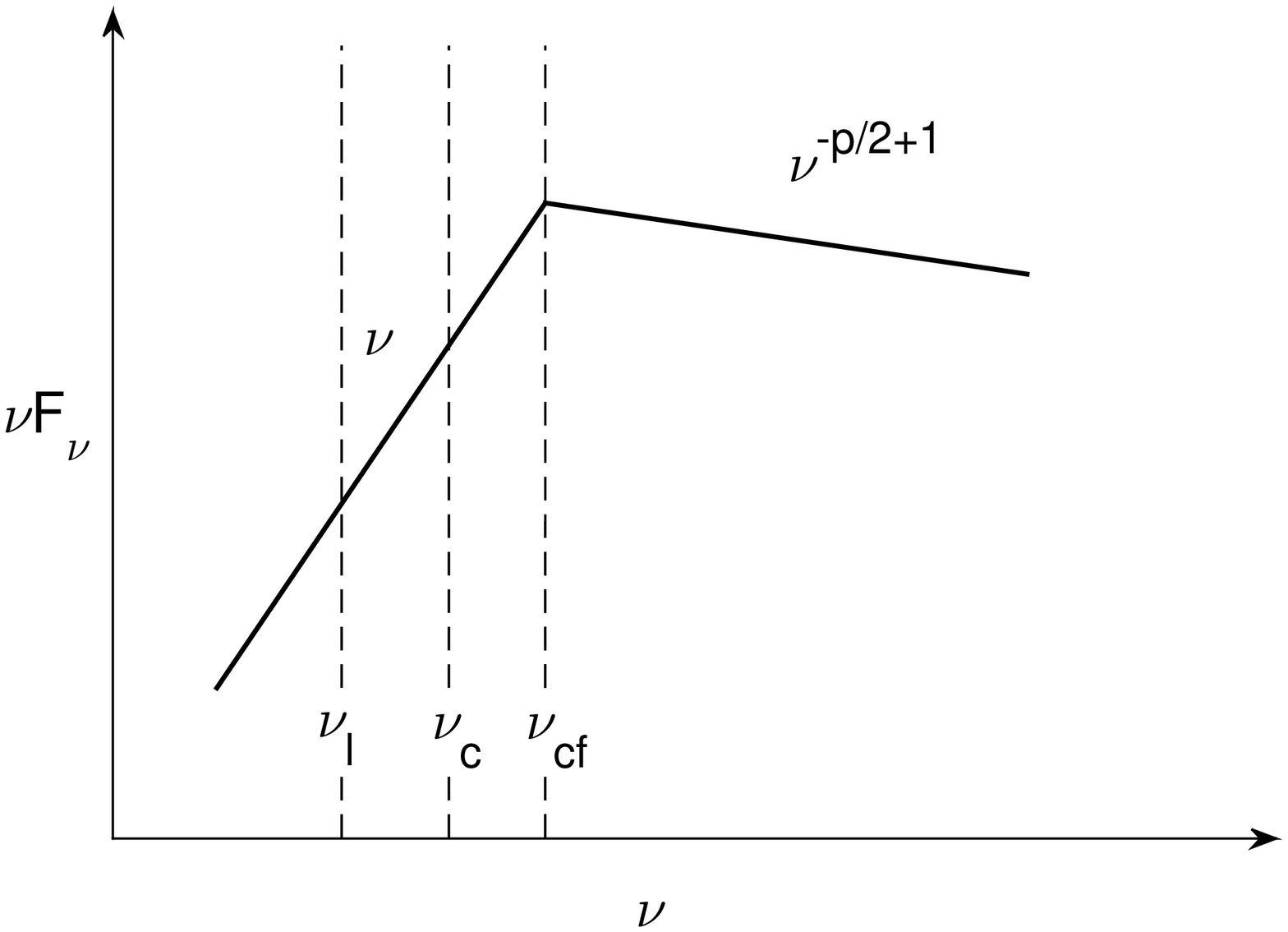}\label{fig: slc1s}}
   
\subfigure[{\it Case (iv)}]{
   \includegraphics[width=8.5cm]{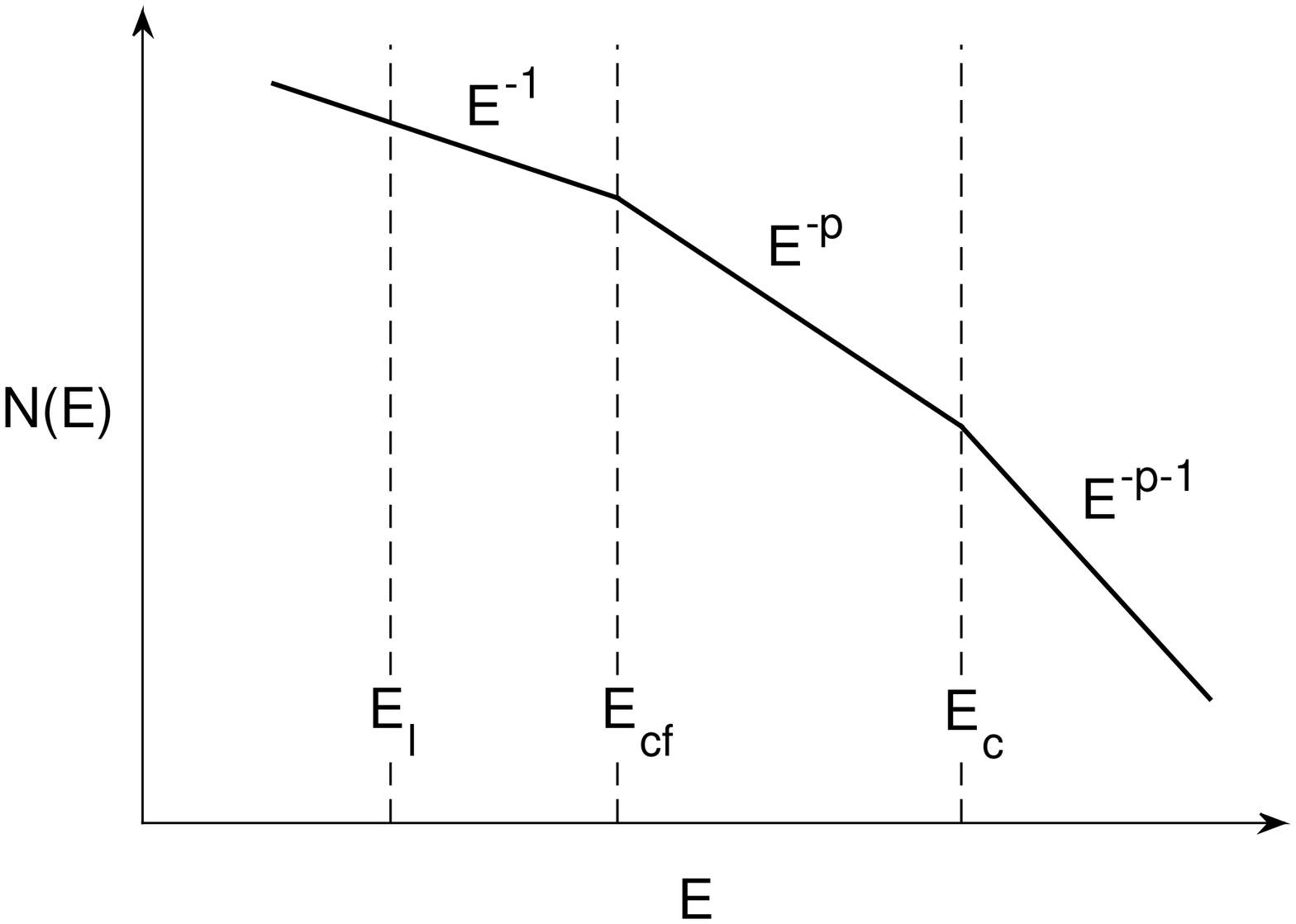}\label{fig: slc2}}
\subfigure[{\it Case (iv)}]{
   \includegraphics[width=8.5cm]{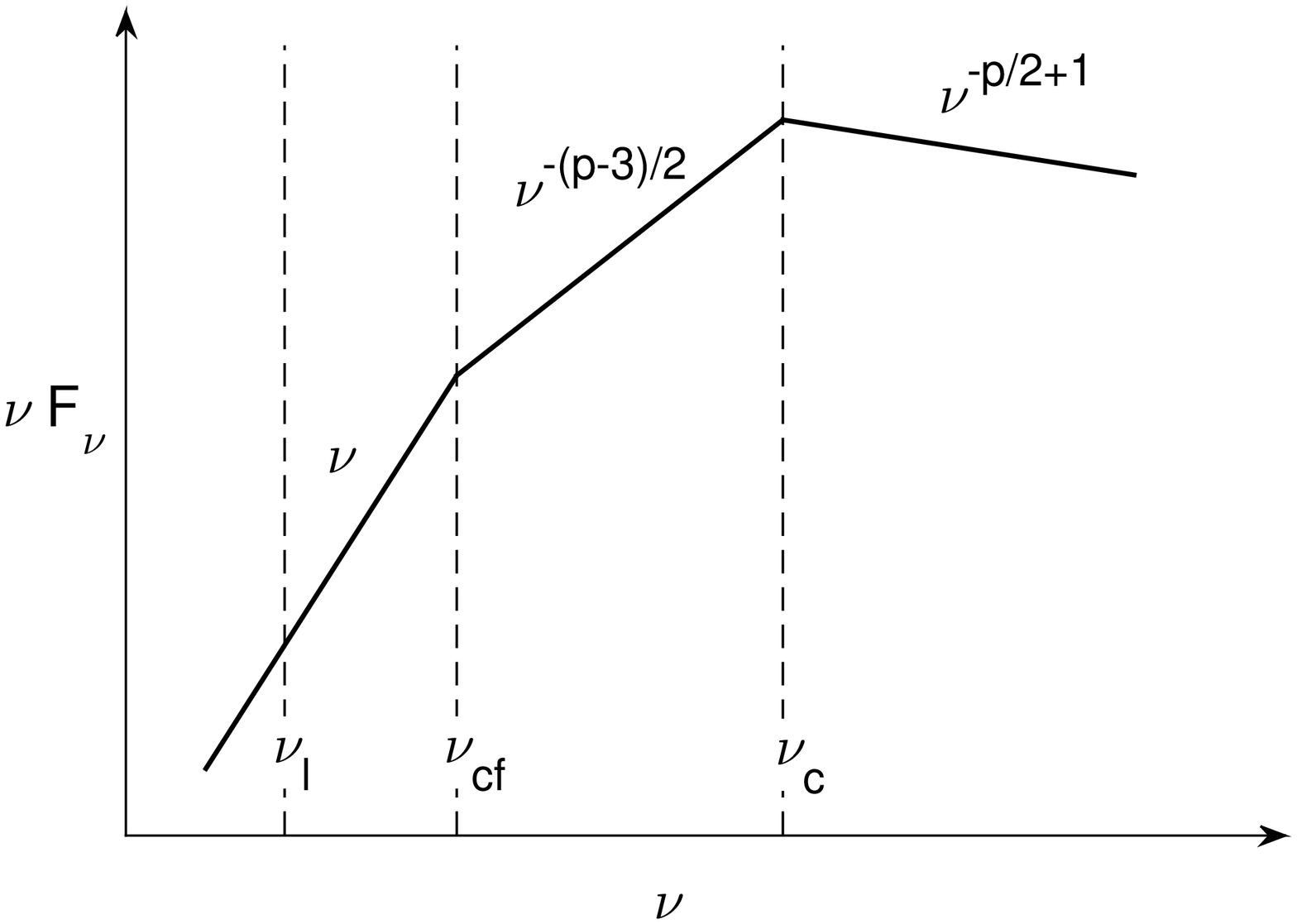}\label{fig: slc2s}}
   
\caption{ The same as Fig. \ref{fig: fast} but for {\it Case (iii)} and {\it Case (iv)}. }
\label{fig: slow}
\end{figure*}

As shown in Fig. \ref{fig: slc1},
$N(E)$ in this situation has the same form as that in {\it Case (i)}. 
The resulting synchrotron spectrum also has the same shape as that in 
Eq. \eqref{eq: fcc1}, which can be seen in Fig. \ref{fig: slc1s}.

{\it Case (iv)}: $ E_l < E_\text{cf} < E_c$

The cooled electrons at $E > E_c$ still follow the spectral shape as in Eq. \eqref{eq: cols}, 
while the electrons in the energy range $E_\text{cf} < E < E_c$ are not subjected to the cooling effect, 
so their original spectral form remains (see Fig. \ref{fig: slc2}).
Then the synchrotron spectrum is given by (Fig. \ref{fig: fac2s}),
\begin{subnumcases}
 {\nu F_\nu=\label{eq: casiv}}
\nu F_{\nu,\text{max}} \propto \nu ,~~~~~~~~~~~~~~~~~~~~~~~~~~~~~~ \nu <\nu_\text{cf}, \\
\nu F_{\nu,\text{max}} \Big(\frac{\nu}{\nu_\text{cf}}\Big)^{-\frac{p-1}{2}} \propto \nu^{-\frac{p-3}{2}}, \nonumber\\
~~~~~~~~~~~~~~~~~~~~~~~~~~~~~~~~~~~~~~~~~~\nu_\text{cf} < \nu < \nu_c, \\
\nu F_{\nu,\text{max}} \Big(\frac{\nu_c}{\nu_\text{cf}}\Big)^{-\frac{p-1}{2}} \Big(\frac{\nu}{\nu_c}\Big)^{-\frac{p}{2}} \propto \nu^{-\frac{p}{2}+1}, \nonumber\\
~~~~~~~~~~~~~~~~~~~~~~~~~~~~~~~~~~~~~~~~~~~~~~~~~~~~\nu > \nu_c,
\end{subnumcases}
where $\nu_c$ corresponds to $E_c$. 
It has a similar form to that in {\it Case (ii)} (Eq. \eqref{eq: fcc2}) except for the intermediate segment.

In all the above cases, we also expect $\nu F_\nu \propto \nu^{4/3}$ at $\nu< \nu_m$ (not shown in Figs. \ref{fig: fast} and \ref{fig: slow}), 
where $\nu_m$ corresponds to $E_m$. 
This low-frequency power-law index is a characteristic feature of synchrotron radiation and is 
independent of the detailed electron energy spectrum 
\citep{Ry79,MR93}.

The above analysis shows that the dominance of the ASA at low energies and synchrotron cooling at high energies 
naturally gives rise to broken power-law electron and synchrotron spectra. 
Moreover, 
depending on the relation between $E_l$, $E_\text{cf}$, and $E_c$, the spectrum can have various forms at intermediate energies (frequencies).

\section{Comparisons between analytical spectral shapes and observed spectra of PWNe}

The synchrotron emission of PWNe is characterized by a broad-band spectrum.
It typically has a flat radio component with $F_\nu \propto \nu^\alpha$,
where $ -0.3 \lesssim \alpha \lesssim  0$,  
and a steep X-ray component with $\alpha \sim -1$
\citep{Che05,Gae06,Rey17}.
We apply the above theoretical model to interpreting the spectral shape of the Mouse PWN (Section \ref{ssec: mou})
and some other PWNe with well-characterized multi-wavelength spectra (Section \ref{ssec: otpwn}). 
Here we aim at explaining the overall spectral shapes,
rather than providing detailed fits for individual sources.

\subsection{Application to the Mouse PWN}\label{ssec: mou}

\begin{figure*}[htbp]
\centering
\subfigure[]{
   \includegraphics[width=8.8cm]{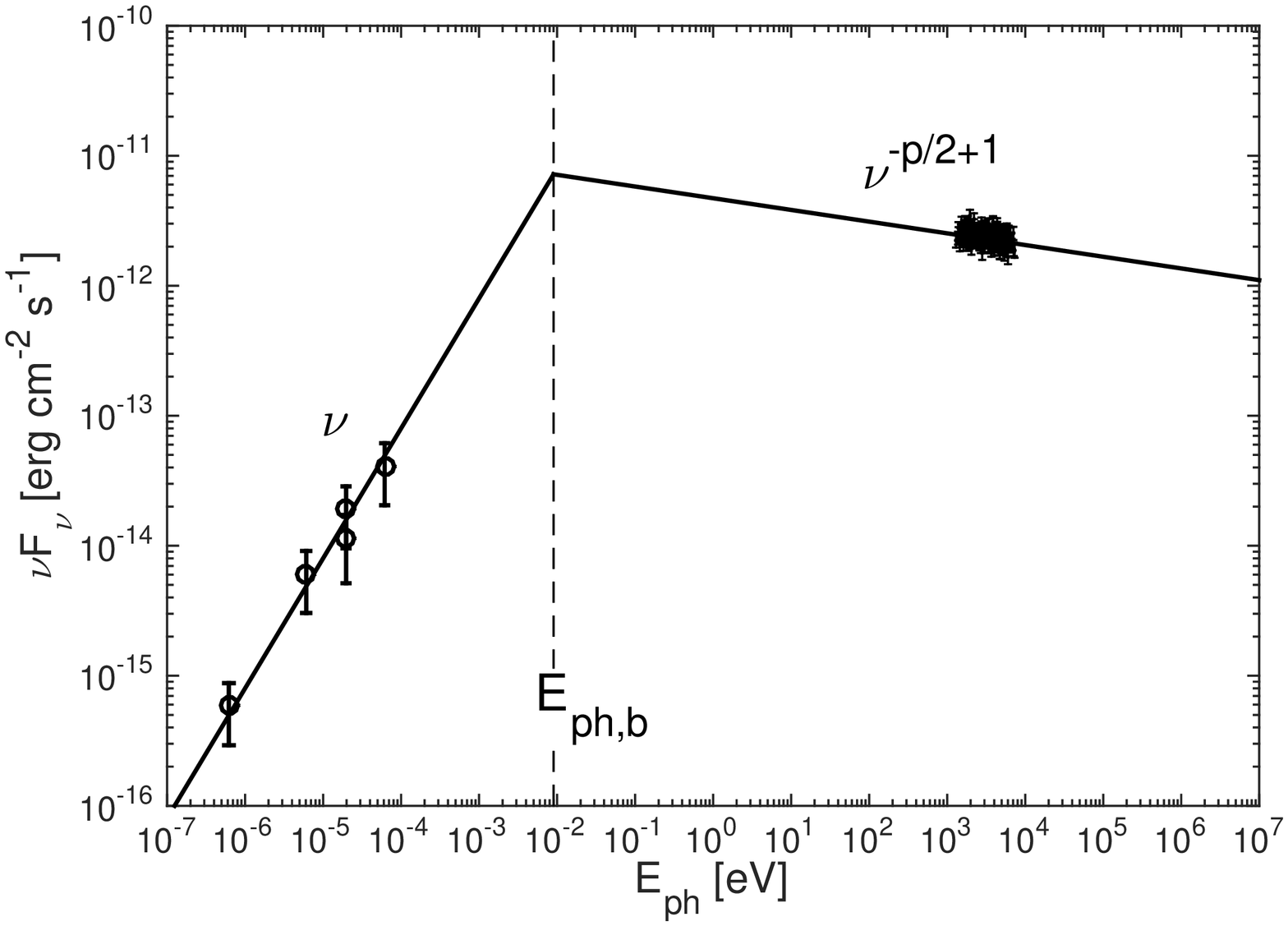}\label{fig:mou}}   
\subfigure[]{
   \includegraphics[width=8.8cm]{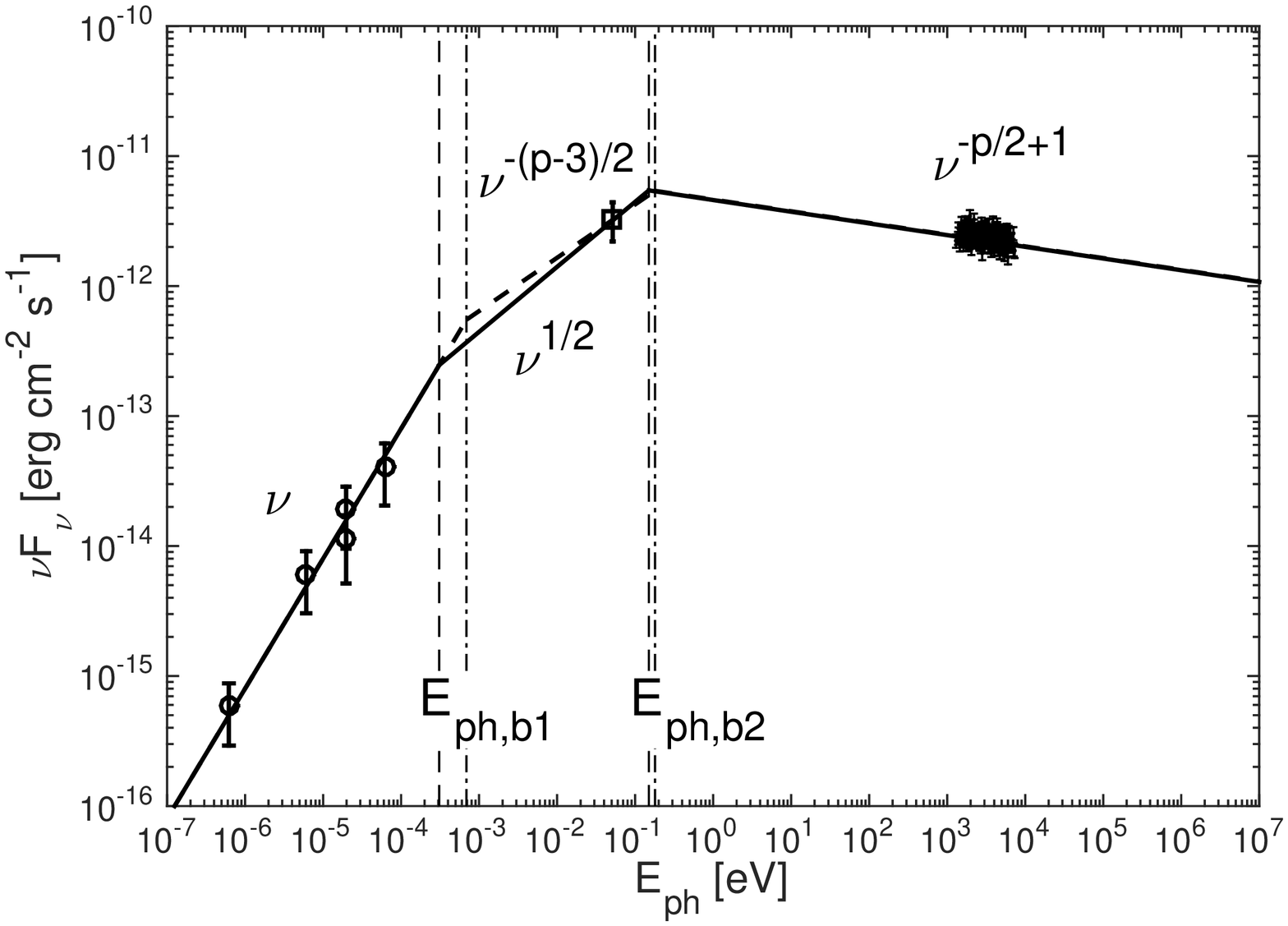}\label{fig:mou2}}   
\caption{Our analytical spectral shapes in comparison with the 
data points of the Mouse PWN taken from 
K18.
Circles, square, and dots represent radio, infrared, and X-ray data, respectively.
The spectral shapes are given by 
Eq. \eqref{eq: fcc1} (solid line) in (a),  
and Eq. \eqref{eq: fcc2} (solid line) and 
Eq. \eqref{eq: casiv} (dashed line) in (b). 
The vertical lines indicate break energies.}
\end{figure*}

Fig. \ref{fig:mou} presents the observed multi-wavelength spectrum of G359.23-0.82, a.k.a., the Mouse PWN powered by PSR J1747-2958.
Over the radio and X-ray data points taken from 
\citet{Klin18} (hereafter K18),
we first assume the spectral shape given in Eq. \eqref{eq: fcc1} corresponding to {\it Case (i)} or {\it Case (iii)} in Section \ref{ssec: case}.
Here we adopt $p=2.18$ 
according to the fit to the X-ray data in 
K18.
We see that the two spectral segments dominated by the ASA and radiation losses
separately match the radio and X-ray data.

The photon energy $E_\text{ph}$ is related to the electron Lorentz factor $\gamma_{e}$ by 
\begin{equation}\label{eq: brene}
    E_\text{ph}   
   = \hbar \frac{e B}{m_e c} \gamma_{e}^2 \Gamma  ,  
\end{equation}
with $\hbar = h/2\pi$, the Planck constant $h$, the electron charge $e$,  
the Lorentz factor $\Gamma (\approx 1)$ of the mildly relativistic bulk flow in the PWN
\citep{Rey17}.
The lowest observed photon energy $E_\text{ph,m} = 6.2\times10^{-7}$~eV provides the upper limit of 
$\gamma_{e,\text{m}} = E_m/(m_e c^2)$,   
\begin{equation}\label{eq: gamina}
      \gamma_{e,\text{m,max}} = 5.2\times10^2 \Big(\frac{\Gamma}{1}\Big)^{-\frac{1}{2}} \Big(\frac{B}{200 ~\mu \text{G}}\Big)^{-\frac{1}{2}} \Big(\frac{E_\text{ph,m}}{6.2 \times10^{-7}~  \text{eV}}\Big)^\frac{1}{2} .
\end{equation}
Here we adopt the equipartition magnetic field $B \sim 200~\mu$G estimated in K18.

The spectral break appears at the photon energy $E_\text{ph,b} \approx 9 \times10^{-3} $ eV.     
It is related to the cutoff electron energy $E_\text{cf} = \gamma_{e,\text{cf}} m_e c^2$ 
and signifies the balance between acceleration and cooling. 
So we have 
\begin{equation}\label{eq: gambon}
    \gamma_{e,\text{cf}} 
    = 6.2\times10^4  \Big(\frac{\Gamma}{1}\Big)^{-\frac{1}{2}} \Big(\frac{B}{200 ~\mu \text{G}}\Big)^{-\frac{1}{2}} \Big(\frac{E_\text{ph,b}}{9 \times10^{-3}~  \text{eV}}\Big)^\frac{1}{2} .
\end{equation}
It gives the upper limit of the minimum Lorentz factor of injected electrons $\gamma_{e,l} = E_l/(m_e c^2)$
in this scenario. 
It also gives the minimum cooling time for the condition $E_c < E_\text{cf}$ to be satisfied
(Eqs. \eqref{eq: betp}, \eqref{eq: eneco}, and \eqref{eq: gambon}),
\begin{equation}\label{eq: mincot}
\begin{aligned}
    t_\text{c,min} &= \frac{1}{E_\text{cf} \beta} = \frac{1}{\gamma_{e,\text{cf}} m_e c^2 \beta}  \\
  &   = 9.9   \Big(\frac{\Gamma}{1}\Big)^{\frac{1}{2}} \Big(\frac{B}{200 ~\mu \text{G}}\Big)^{-\frac{3}{2}} \Big(\frac{E_\text{ph,b}}{9 \times10^{-3}~  \text{eV}}\Big)^{-\frac{1}{2}} ~\text{kyr}. 
\end{aligned}
\end{equation}
Besides, the maximum cooling time $t_\text{c,max}$ is given by the 
spin-down age $t_\text{sd} = 25.5$~kyr of the pulsar 
(K18).
It corresponds to the electron Lorentz factor
(Eqs. \eqref{eq: betp}, \eqref{eq: eneco}, and \eqref{eq: gamina}),
\begin{equation}
\begin{aligned}
     \gamma_{e,\text{sd}} &  =  \frac{1}{m_e c^2 \beta t_\text{sd}} \\
                                        & =   2.4\times10^4               \Big(\frac{B}{200 ~\mu \text{G}}\Big)^{-2} \Big(\frac{t_\text{sd}}{ 25.5~\text{kyr}}\Big)^{-1}        
                                       &    > \gamma_{e,\text{m,max}},
\end{aligned}
\end{equation}
implying that the electrons in the Mouse PWN are in the slow cooling regime with $E_{c,\text{min}} > E_{m,\text{max}}$
(see also K18 for the analysis of the cooling regime).

Moreover, $t_\text{c,min}$ in Eq. \eqref{eq: mincot} also gives the 
acceleration timescale of the ASA (Eq. \eqref{eq: cutofe}),
\begin{equation}\label{eq: accts}
     \tau_\text{acc} = a_A^{-1} = t_\text{c,min}. 
\end{equation}
We note that $\tau_\text{acc}$ in {\it Case (i)} or {\it Case (iii)} is shorter than $t_c$.

Including the tentative infrared point (see Fig. \ref{fig:mou2} and K18) 
\footnote{We caution that the connection of this infrared emission to the Mouse PWN needs further examination 
by high-resolution infrared observations
(K18).}
requires the spectral shape given by Eq. \eqref{eq: fcc2} in {\it Case (ii)} or 
Eq. \eqref{eq: casiv} in {\it Case (iv)}. 
The spectral breaks at lower and higher energies are 
$E_\text{ph,b1} \approx 3.1\times10^{-4}$~eV, 
$E_\text{ph,b2}\approx 0.15$~eV in {\it Case (ii)}, 
and 
$E_\text{ph,b1} \approx 6.9\times10^{-4}$~eV,
$E_\text{ph,b2}\approx 0.18$~eV in {\it Case (iv)}. 
Similar to the above analysis, these break energies can be used to infer the characteristic $\gamma_e$,
$t_\text{c}$, and $\tau_\text{acc}$
(Eqs. \eqref{eq: brene}, \eqref{eq: mincot}, and \eqref{eq: accts}), as shown in Table \ref{tab: moug}.
We see that since the minimum $t_\text{c}$ required in {\it Case (ii)} is much larger than $t_\text{sd}$ of the pulsar, 
{\it Case (ii)} is disfavored.

\begin{table*}[htbp]
\renewcommand\arraystretch{1.7}
\centering
\begin{threeparttable}
\caption[]{Characteristic parameters inferred from the spectral breaks of the Mouse PWN}\label{tab: moug} 
  \begin{tabular}{c|c|c|c}
     \toprule
         $E_\text{ph}$                                        & $E_\text{ph,m}$                                                                  &  $E_\text{ph,b}$ or $E_\text{ph,b1}$ & $E_\text{ph,b2}$    \\
               \hline
           {\it Case (i)} or {\it Case (iii)}    & \multirow{3}{*}{$\gamma_{e,\text{m}} \approx 5.2\times10^2$}     &  $\gamma_{e,\text{cf}} = \gamma_{e,l,\text{max}} \approx 6.2\times10^4$, $\tau_\text{acc} = t_\text{c,min}\approx 9.9$~kyr    &   ---  \\ 
           {\it Case (ii)}                             &                                                                                                        &  $\gamma_{e,\text{cf}} \approx1.1\times10^4$,   $\tau_\text{acc} = t_\text{c,min} \approx 53.5$~kyr  &  $\gamma_{e,l} \approx 2.5\times10^5$ \\
           {\it Case (iv)}                            &                                                                                                        &  $\gamma_{e,\text{cf}}= \gamma_{e,l,\text{max}}\approx 1.7\times10^4$,$\tau_\text{acc} \approx 35.8$~kyr  &  $t_\text{c} \approx 2.2$~kyr \\
     \bottomrule
    \end{tabular}
  \end{threeparttable}
\end{table*}

The above comparisons show that 
our analytical spectral shapes agree well with the observed multi-wavelength spectrum of the Mouse PWN. 
The ASA and synchrotron cooling in the PWN naturally account for the flatness and steepness
of the radio and X-ray spectra ($F_\nu$), respectively. 
More reliable infrared measurements with e.g., {\sl HST}, {\sl JWST}, and {\sl ALMA} 
are needed to determine the detailed spectral shape at intermediate energies
and to distinguish between {\it Case (i)} (or {\it Case (iii)}) and {\it Case (iv)}.

\subsection{Application to some other PWNe}\label{ssec: otpwn}

We next compare our analytical spectral shapes with the synchrotron spectra of some other PWNe 
to further examine the applicability of our model.

\begin{figure*}[htbp]
\centering   

\subfigure[]{
   \includegraphics[width=8.5cm]{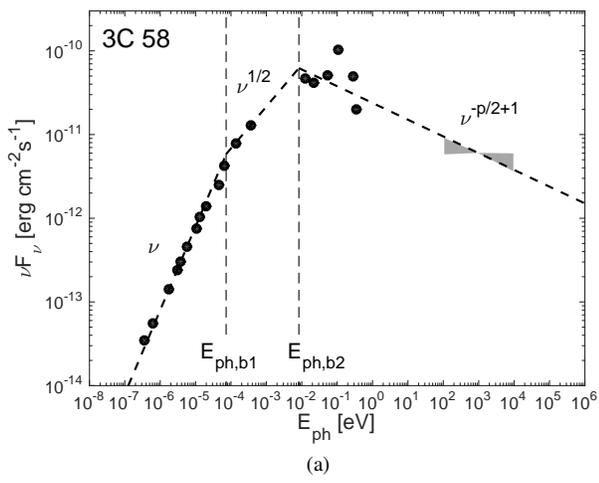}\label{fig: 3c}}  
\subfigure[]{
   \includegraphics[width=8.5cm]{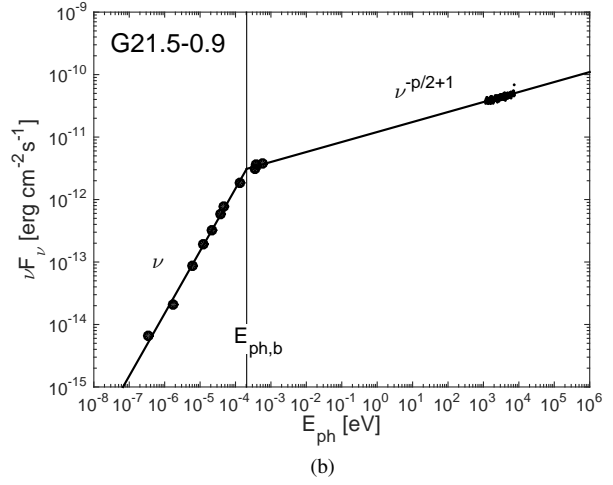}\label{fig: g21}}    
   
\subfigure[]{
   \includegraphics[width=8.5cm]{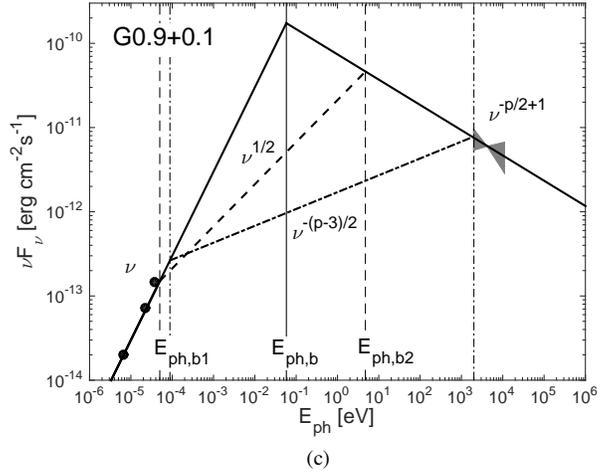}\label{fig: g09}}      

\subfigure[]{
   \includegraphics[width=8.5cm]{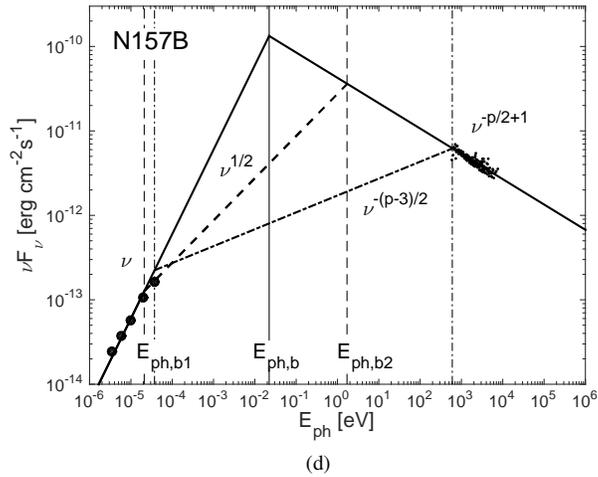}\label{fig: n157}}      
    
\caption{ Our analytical spectral shapes (solid line: Eq. \eqref{eq: fcc1}, dashed line: Eq. \eqref{eq: fcc2}, dash-dotted line: Eq. \eqref{eq: casiv})
in comparison with the observed synchrotron spectra of PWNe.
Vertical lines indicate break energies. }
\label{fig: colle}
\end{figure*}

In Fig. \ref{fig: colle}, we display the synchrotron spectra of 
3C 58
\citep{Sla08},
G21.5-0.9
\citep{Sal89},
G0.9+0.1
\citep{Tan11}, 
and 
N157B
\citep{Zhu18}.
We also include the infrared data for 3C 58, 
but we caution that the observed values are only upper limits 
\citep{Sla08}.
The X-ray data for G21.5-0.9 and N157B 
were obtained from {\sl Chandra} ObsIDs 
2873, 3699 3700, 5158, 5159, 5165, 5166, 6070, 6071, 6740, 6741, 8371, 8372 (G21.5--0.9), and 
2783 (N157B).  
The data were reprocessed using the {\sl Chandra} Interactive Analysis of Observations (CIAO) software version 4.9 and the {\sl CXO} Calibration Data Base (CALDB) version 4.7.7.  All observations were reprocessed with {\tt chandra\_repro} to apply the latest calibrations.  We extracted the spectra using {\tt specextract} and fitted the power-law spectra using the {\tt tbabs} model of XSPEC (v12.9.1p, which uses absorption cross-sections from 
\citet{Wil00}).  
For G21.5--0.9 we use a circular $2\farcm5$ aperture centered on the pulsar, and excluding the bright point source to the South. 
For N157B, we use a 53$''$ $\times$ 24$''$ extraction region approximately coincident with the brightest radio contour shown in Figure 3 of 
\citet{Ch06}
and free of thermal supernova remnant (SNR) filaments.  
In both cases we use background regions located outside the extent of the PWNe/SNRs, 
and restricted the spectral fits to the 0.5--8 keV range.  
We verified that our results are not noticeably contaminated by thermal SNR emission by restricting the spectral fits to energies above 2 keV and re-fitting; in both cases the resulting power-law slopes were virtually identical.
Our fits to the X-ray data imply $p=1.68$ for G21.5-0.9
and $p=2.6$ for N157B. 
For 3C 58 and G0.9+0.1, we use $p=2.4$ and $p=2.6$ according to the fits to their X-ray data 
\citep{Sla08,Tan11}.

For 3C 58 and G21.5-0.9, as indicated by the data, 
we apply the spectral shape in Eq. \eqref{eq: fcc2} ({\it Case (ii)}) and that in Eq. \eqref{eq: fcc1} ({\it Case (i)} or {\it Case (iii)}), respectively. 
For G0.9+0.1 and N157B, the measurements are insufficient to constrain the detailed spectral form in the intermediate energy range, 
so we present all possible spectral shapes for comparison with future observations. 
In all cases, 
it is clear that the electrons in the PWN should undergo the stochastic acceleration to 
account for the radio spectrum. 
The general agreement between our analytical spectral shapes and observed spectra implies 
the ASA and synchrotron cooling together as a common 
origin of broad-band synchrotron spectra of PWNe.

\begin{figure}[htbp]
\centering
\includegraphics[width=9cm]{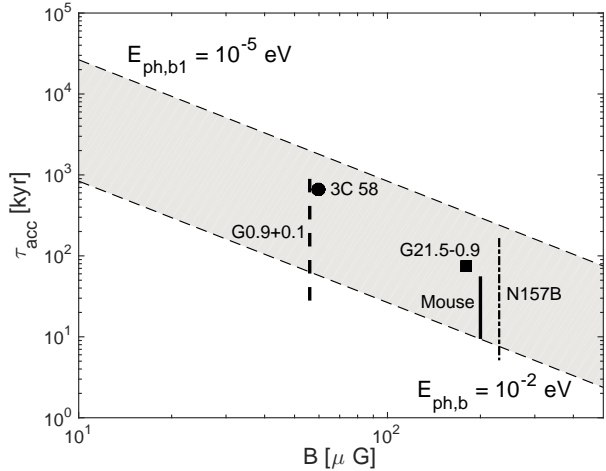}
\caption{ $\tau_\text{acc}$ vs. $B_\text{eq}$ of the PWNe. 
The shaded region is bounded by Eq. \eqref{eq: parang}. }
\label{fig: sum}
\end{figure}

The spectral break between radio and X-ray bands or 
radio and infrared bands is determined by $\tau_\text{acc}$ of the ASA. 
When $\tau_\text{acc}$ is sufficiently short for $E_\text{cf}$ to be larger than both $E_c$ and $E_l$, 
the synchrotron spectrum of a PWN falls in {\it Case (i)} or {\it Case (iii)}. 
Then we can rewrite Eq. \eqref{eq: mincot} as 
\begin{equation}\label{eq: cobss}
    \Big(\frac{\tau_\text{acc} }{1~\text{kyr}}\Big) \Big(\frac{B}{100 ~\mu \text{G}}\Big)^{\frac{3}{2}}
     = 26.5   \Big(\frac{\Gamma}{1}\Big)^{\frac{1}{2}}  \Big(\frac{E_\text{ph,b}}{ 10^{-2}~  \text{eV}}\Big)^{-\frac{1}{2}} , 
\end{equation}
and obtain a constraint on $\tau_\text{acc}$ and $B$ in the PWN. 
When $\tau_\text{acc}$ is relatively long with $E_\text{cf} < E_l$ or $E_\text{cf} < E_c$, 
the PWN is in {\it Case (ii)} or {\it Case (iv)}. 
Accordingly, we have 
\begin{equation}\label{eq: coattw}
   \Big( \frac{\tau_\text{acc} }{1~\text{kyr}} \Big) \Big(\frac{B}{100 ~\mu \text{G}}\Big)^{\frac{3}{2}}
     = 837.0   \Big(\frac{\Gamma}{1}\Big)^{\frac{1}{2}}  \Big(\frac{E_\text{ph,b1}}{ 10^{-5}~  \text{eV}}\Big)^{-\frac{1}{2}}  .
\end{equation}
If the spectral shape in the infrared band cannot be observationally determined 
(e.g., G0.9+0.1, N157B),
we can still set a constraint on the range of $\tau_\text{acc}$ and $B$,
\begin{equation}\label{eq: parang}
\begin{aligned}
26.5   \Big(\frac{\Gamma}{1}\Big)^{\frac{1}{2}}  \Big(\frac{E_\text{ph,b}}{ 10^{-2}~  \text{eV}}\Big)^{-\frac{1}{2}} & \leq
\Big( \frac{\tau_\text{acc} }{1~\text{kyr}} \Big) \Big(\frac{B}{100 ~\mu \text{G}}\Big)^{\frac{3}{2}} \\
 & \leq 837.0   \Big(\frac{\Gamma}{1}\Big)^{\frac{1}{2}}  \Big(\frac{E_\text{ph,b1}}{ 10^{-5}~  \text{eV}}\Big)^{-\frac{1}{2}}  .
\end{aligned}
\end{equation}
In Table \ref{tab: para}, we list the values of $E_\text{ph,b}$ and $E_\text{ph,b1}$ of the above PWNe.
For G0.9+0.1 and N157B, we can only provide the lower limit of $E_\text{ph,b1}$ due to the insufficient data points. 
By using the estimated magnetic field strength $B_\text{eq}$ under the assumption of equipartition between particle and magnetic energies 
(see Table \ref{tab: para}), 
in Fig. \ref{fig: sum}
we present the corresponding $\tau_\text{acc}$ (Eqs. \eqref{eq: cobss} and \eqref{eq: coattw}) or the range of $\tau_\text{acc}$ (Eq. \eqref{eq: parang})
inferred from the spectral breaks of the PWNe. 
Empirically,
we see that $\tau_\text{acc}$ of different PWNe are basically within the range confined by Eq. \eqref{eq: parang}.

\begin{table*}[htbp]
\centering
\begin{threeparttable}
\caption[]{Parameters of the PWNe}\label{tab: para} 
  \begin{tabular}{cccccc}
     \toprule
    PWNe                      & Mouse                             &  3C 58                  &   G21.5-0.9                  & G0.9+0.1                         &  N157B                        \\
      \hline
  $E_\text{ph,b}$ [eV]   &  $9.0\times10^{-3}$        &   ---                       &    $2.1\times10^{-4}$   & $5.8\times10^{-2}$    & $2.2\times10^{-2}$    \\
  $E_\text{ph,b1,min}$ [eV]  & $3.1\times10^{-4}$ & $7.3\times10^{-5}$  & ---                             & $5.0\times10^{-5}$       & $2.1\times10^{-5}$       \\
 $B_\text{eq}$ [$\mu$G]  &  $200$\tnote{(1)}        & $60$\tnote{(2)}         & $180$\tnote{(3)}       & $56$\tnote{(4)}           &  $230$\tnote{(5)}                 \\
    \bottomrule
    \end{tabular}
 \begin{tablenotes}
      \small
      \item[(1)]   K18.
      \item[(2)]   \citet{Rey88}.
      \item[(3)]   \citet{Saf01}.
      \item[(4)]   \citet{Dub08}.
      \item[(5)]   \citet{Wan01}
    \end{tablenotes}
 \end{threeparttable}
\end{table*}

\section{Discussion}

Turbulence is believed to be present in PWNe, 
but its origin and properties are unclear. 
Besides the magnetic reconnection under consideration here, 
other possible sources of turbulence include, e.g., 
the Rayleigh-Taylor instability induced by the interaction between the PWN and the supernova remnant 
\citep{Che75,Por14},
the dynamical instability of the PW's oscillating magnetic field 
\citep{Zrak16},
the Weibel instability
\citep{Wei59,Sir11},
the large-scale shear flows in the PWN 
\citep{Kom04}. 
For bow-shock PWNe, an additional source of turbulence can be 
the Richtmyer-Meshkov instability induced by their interaction with interstellar density fluctuations 
\citep{Giac_Jok2007}.
As a more general mechanism to generate turbulence, 
the magnetic reconnection is dynamically coupled with turbulence and 
naturally converts magnetic energy to turbulent energy. 
The generated turbulence allows efficient dissipation of magnetic energy through turbulent reconnection in the PW, 
and further ASA of the electrons in the PWN.

Non-adiabatic stochastic acceleration including the gyroresonance and transit time damping (TTD)
can also result in a hard electron spectrum 
\citep{Mel69}.
As is known, 
the gyroresonance with Alfv\'{e}n modes is inefficient due to the turbulence anisotropy 
\citep{Chan00,YL02},
while TTD with compressive fast and slow modes can be alternative stochastic acceleration mechanisms
\citep{YL08,Xult}.
However, only a small fraction of turbulent energy is contained in compressive modes in compressible MHD turbulence 
\citep{CL02_PRL,KowL10}.
Moreover, fast modes are usually subjected to severe plasma damping effects 
\citep{YL04},
and TTD with slow modes also suffers the turbulence anisotropy 
\citep{Xult}.
We note that the turbulence anisotropy is scale-dependent
\citep{CL03}. 
Even at a low magnetization in outer regions of PWNe, the turbulent eddies at the small resonant scale are still highly anisotropic.
Therefore, when the pitch-angle scattering and non-adiabatic stochastic acceleration are inefficient, 
the ASA can account for the stochastic acceleration in PWNe. 
Since the largest eddies are most effective in ASA, its efficiency does not depend on the turbulence spectral slope.
We find numerical evidence for the ASA in both MHD simulations 
\citep{Lyn12}
and kinetic particle-in-cell simulations
\citep{Dah17}.
Obviously, in a situation where the pitch-angle scattering is efficient and thus the adiabatic invariance is violated, 
the ASA is suppressed.

The ASA occurs when the Larmor radius $r_L$ of electrons is much smaller than the characteristic length scale of turbulent magnetic fields. 
The largest $r_L$ of the stochastically accelerated electrons corresponding to $E_\text{ph,b}$ is (Eq. \eqref{eq: gambon})
\begin{equation}
\begin{aligned}
     r_L  &= \frac{\gamma_{e,\text{cf}} m_e c^2}{e B} \\
            & = 1.6\times10^{12} \Big(\frac{\Gamma}{1}\Big)^{-\frac{1}{2}} \Big(\frac{B}{100 ~\mu \text{G}}\Big)^{-\frac{3}{2}} \Big(\frac{E_\text{ph,b}}{10^{-2}~  \text{eV}}\Big)^\frac{1}{2} ~\text{cm} .
\end{aligned}
\end{equation}
It is indeed small compared with 
the characteristic turbulence scale of the order of $0.1$ pc
suggested by polarization observations of PWNe   
(e.g., \citealt{Yus05,Mor13,Ma16}).
Given the turbulence scale $l_\text{tur}$ and $\tau_\text{acc}$, we can also estimate the turbulent speed at $l_\text{tur}$
(Eqs. \eqref{eq: a2}, \eqref{eq: mincot}, and \eqref{eq: accts})
\begin{equation}
\begin{aligned}
        u_\text{tur} & =  \xi^{-1} l_\text{tur}   \tau_\text{acc}^{-1}            \\
                          & =  3.7\Big(\frac{ \xi}{1}\Big)^{-1}     \Big(\frac{\Gamma}{1}\Big)^{-\frac{1}{2}} \Big(\frac{B}{100 ~\mu \text{G}}\Big)^{\frac{3}{2}} \Big(\frac{E_\text{ph,b}}{ 10^{-2}~  \text{eV}}\Big)^{\frac{1}{2}}  \\
                      &~~~~~~    \Big(\frac{l_\text{tur}}{0.1~\text{pc}}\Big)~\text{km s$^{-1}$}.
\end{aligned}
\end{equation}
If we assume that the interstellar turbulence driven by supernova explosions has $L_i \sim 100$~pc and the 
turbulent speed $V_L \sim 10$~km s$^{-1}$ at $L_i$
\citep{Kap70},
then according to the Kolmogorov scaling
\citep{Armstrong95,CheL10}, 
the turbulent speed in the interstellar medium at $0.1$ pc is $v_l = V_L (l/L_i)^{1/3} \sim 1$~km s$^{-1}$. 
It implies that the turbulence level in a PWN can be comparable to the surrounding interstellar turbulence.

We have analyzed the synchrotron spectra of both 
bow-shock PWN
(Mouse)
and PWNe supernova remnants
(3C 58, G21.5-0.9, G0.9+0.1, N157B). 
Despite their possibly different interaction with ambient media, 
all PWNe exhibit broken power-law spectra with a flat radio component ($F_\nu$) and a steeper X-ray component. 
This common feature implies that both stochastic acceleration and synchrotron cooling of electrons generally exist 
in different types of PWNe.
Relativistic MHD simulations of PWNe also support that radio emitting particles undergo a local re-acceleration process 
in the nebula by the interaction with turbulence 
\citep{Olm14,Olm16}.

The ASA is a promising mechanism of particle re-acceleration to explain the flat $F_\nu$ spectra of different sources. 
Besides gamma-ray bursts discussed in earlier studies 
\citep{XZg17,Xu18}
and PWNe considered here, 
it can also be applicable to e.g. 
radio galaxies,
blazars,
which will be investigated in our future work.

\section{Summary}

Recent advances in the theoretical understanding of MHD turbulence bring us new insights 
in basic physical processes related to PWNe. 

We suggested that the magnetic reconnection in the PW inevitably falls in the turbulent reconnection regime 
and thus is efficient in converting the magnetic energy to the particle energy. 

We focused on the re-acceleration of particles in the PWN through the mechanism ASA. 
The ASA acts to flatten the injected energy spectrum of electrons.
It results in a flat radio spectrum 
(in $F_{\nu}$) of the photons radiated downstream of the termination shock.
The steeper X-ray spectrum can be simply attributed to synchrotron cooling, which is dominant at high energies. 
At intermediate energies, the detailed spectral shape depends on the acceleration timescale of the ASA.

Our analytical spectral shapes incorporating both the ASA and synchrotron cooling 
generally agree well with the observed broad-band synchrotron spectra of PWNe. 
The spectral breaks can be used to constrain the acceleration timescale of the ASA and the magnetic field strength in the PWN (see Fig. \ref{fig: sum}). 
They can also provide valuable information on the cooling time and injected electron energy 
(see Table \ref{tab: moug}).

\section*{Acknowledgement}

SX acknowledges the support for Program number HST-HF2-51400.001-A provided by NASA through a grant from the Space Telescope Science Institute, which is operated by the Association of Universities for Research in Astronomy, Incorporated, under NASA contract NAS5-26555.
Support for NK and OK was provided by the National
Aeronautics and Space Administration through
Chandra
Award Number G03-14082 issued by the
Chandra
X-ray
Observatory Center, which is operated by the Smithsonian Astrophysical Observatory for and on behalf of
the National Aeronautics and Space Administration under contract NAS8-03060. 
BZ acknowledges the 
NASA grant NNX15AK85G.

\bibliographystyle{apj.bst}
\bibliography{xu}

\end{document}